%% file: main.tex
\documentclass[12pt]{article}
\input{preamble}

\date{}
\title{Perfect Fractional Matchings in Bipartite Graphs Via Proportional Allocations\thanks{This material is based upon work supported in part by the Air Force Office of Scientific Research under award number FA9550-23-1-0031. D.~Hathcock supported by NSF Graduate Research Fellowship grant DGE-2140739.
}}

\author{Daniel Hathcock\thanks{Corresponding Author: {\tt dhathcoc@alumni.cmu.edu}, +1 303-834-2557, Carnegie Mellon University, 5000 Forbes Ave, Pittsburgh, PA 15213, USA.}
\and 
R. Ravi\thanks{{\tt ravi@andrew.cmu.edu}, Carnegie Mellon University, USA.}
}

\begin{document}

\maketitle

\begin{abstract}
Given a bipartite graph that has a perfect matching, a prefect proportional allocation is an assignment of positive weights to the nodes of the right partition so that every left node is fractionally assigned to its neighbors in proportion to their weights, and these assignments define a fractional perfect matching. We prove that a bipartite graph has a perfect proportional allocation if and only if it is matching covered, by using a classical result on matrix scaling. We also present an extension of this result to provide a simple allocation strategy in non-matching covered bipartite graphs.
\end{abstract}


\pagenumbering{gobble}

\pagenumbering{arabic}
\setcounter{page}{1}

\section{Introduction}

Online resource allocation problems arise in a variety of settings, such as allocating impressions in display advertising and assigning arriving supply to warehouses in inventory allocation. In such applications, the arriving items to be allocated are very small compared to the space available to allocate them to (budgets in advertising, warehouse capacity in inventory storage). Moreover, these allocations must be carried out at a massive scale in which the number or speed of arriving items is too large for the typical centralized approaches (the entire network may be too large to fit into memory). Hence, there is a need for simple allocation strategies that can be implemented locally for each arriving item without knowledge of the remainder of the network. One such strategy is called \emph{proportional allocation}, studied extensively by Agrawal, Mirrokni, and Zadimoghaddam~\cite{AMZ18}. 

Classic models of optimal allocations involve building a bipartite graph with one side modeling the items to be allocated and the other modeling the objects to which they are assigned. In this setting, feasible allocations are represented by matchings that assign the items to objects subject to their capacity constraints. Formally, consider the bipartite matching problem in the graph $G = (I \cup J, E)$. The nodes $i \in I$ have an integer \emph{supply} $S_i \in \Z_{\geq 1}$, and the nodes $j \in J$ have integer \emph{capacities} $C_j \in \Z_{\geq 1}$. An allocation is a (possibly fractional) assignment of the supply of nodes in $I$ to nodes in $J$ so that for each $j \in J$, no more than $C_j$ units of supply are assigned to $j$.

In many online allocation problems, since it is typically assumed that the supply of a single arriving item is small compared to the capacities, we are content to find a fractional matching (see the ``small bids'' assumptions in e.g., \cite{MehtaSVV07,FeldmanKMMP09}). To address the need for simple-to-implement allocation strategies that can be computed with only local information in the network, Agrawal, Mirrokni, and Zadimoghaddam proposed proportional allocation~\cite{AMZ18}, in which every node $j \in J$ is given a weight $\alpha_j > 0$. Then each node $i \in I$ is allocated to its neighbors in $G$ proportionally according to their weights. That is, we get the fractional matching $\{x_{ij}\}$ defined by
\[
    x_{ij} := \frac{S_i \cdot \alpha_j}{\sum_{j' \sim i} \alpha_{j'}}
\]
for each $j \sim i$. 

For any given choice of weights, the resulting fractional matching may or may not respect the capacity constraints of the right hand nodes. Therefore, the value of a proportional allocation is defined as the sum of the allocations on each $j \in J$ up to their capacity. Formally,
\[
    \val(x) = \sum_{j \in J} \min\left(C_j, \sum_{i \sim j} x_{ij}\right).
\]
We want a proportional allocation whose value is close or equal to the value of the maximum matching in the bipartite graph (which is clearly an upper bound on the highest possible value). We refer to the value of the maximum matching as $\OPT$. 

Surprisingly, Agrawal, Mirrokni, and Zadimoghaddam~\cite{AMZ18} show that there is always a selection of weights that gives a proportional allocation whose value can be made arbitrarily close (but never equal) to the maximum matching in any bipartite graph. 

\begin{theorem}[\cite{AMZ18}]
For any instance of bipartite matching and any $\eps > 0$, there is an iterative algorithm that finds a selection of weights $\{\alpha_j\}_{j \in J}$ such that the resulting proportional allocation $x$ has value $\val(x) \geq (1-\eps) \OPT$. The algorithm takes $O(\frac{\log (n/\eps)}{\eps^2})$ rounds each running in time $O(m)$, where $n$ and $m$ denote the number of nodes and edges in the graph, respectively.
\end{theorem}

Their result raises a natural question. We refer to a proportional allocation of value $\OPT$ as a \emph{perfect proportional allocation}.
\begin{quote}\centering
    \textit{Does there always exist a choice of weights $\{\alpha_j\}_{j \in J}$ giving a perfect proportional allocation?}
\end{quote}
Unfortunately, the answer in general is ``no''. An instance that demonstrates this is a path on 3 edges with unit supply and capacities. The size of the maximum matching is 2, however, no proportional allocation can get a value of 2. This raises two further questions. 
\begin{quote}\centering
    \textit{For which instances does there exist a choice of weights $\{\alpha_j\}_{j \in J}$ giving a perfect proportional allocation?}
\end{quote}
and 
\begin{quote}\centering
    \textit{Is there an alternate allocation strategy that uses only local information in the network and always gives an allocation of value $\OPT$?}
\end{quote}
In this paper, we use a classical result on matrix scaling to characterize the graphs that admit a perfect proportional allocation, answering the first question, and use this to formulate a new simple strategy that answers the second question in the affirmative, which we call \emph{piecewise proportional allocation}.

\subsection{Related Work}

There is extensive literature, starting with the work of Karp, Vazirani, and Vazirani~\cite{KVV90}, related to finding approximately maximum matchings in bipartite graphs when the nodes on one side arrive online (see \cite{HTW24} for a survey on the topic), and extensions to the Display Ads and AdWords problems (\cite{MehtaSVV07, FeldmanKMMP09, DevenurH09, VeeVS10, GM16, DevanurJSW19}, to list a few). In contrast to these online bipartite matching works, the setting of our paper deals with finding perfect matchings, motivated by an online setting.

Agrawal, Mirrokni, and Zadimoghaddam~\cite{AMZ18} first studied proportional allocations in the context of algorithmic computation of weights that give an approximately optimal allocation. Lavastida et al.~\cite{LMRX21} revisited the topic, giving a simplified algorithm with similar approximation guarantees. However, neither work addressed whether their algorithms would converge to exact optimality.

Matrix scaling and the related Sinkhorn algorithm have a rich history, as surveyed by Idel~\cite{idel2016review}. Rothblum and Schneider~\cite{RS89} and later Hayashi, Hirai, and Sakabe~\cite{HHS24} give precise characterizations of when a matrix is scalable (See Section~\ref{sec:ppa} for a precise definition). 

\section{Preliminaries}

\paragraph{Notation} As above, we will denote an instance of the allocation problem with a bipartite graph $G = (I \cup J, E)$ with supplies $S_i$ for nodes $i \in I$ and capacities $C_j$ for nodes $j \in J$. We use standard graph notation, with $N(i)$ and $N(X)$ denoting the set of neighbors of a node or node set, respectively, and $i \sim j$ meaning that there is an edge connecting $i$ and $j$ (i.e., $i \in N(j)$). 

For a set of nodes $X \subseteq I$, denote by $S_X$ the total supply: $S_X := \sum_{i \in X} S_i$. Similarly, for a set $Y \subseteq J$, denote by $C_Y$ the total capacity: $C_Y := \sum_{j \in Y} C_j$. Also, for a given allocation $x$, we will denote the amount of value allocated to a node $j$ as $\alloc(j) := \sum_{i \sim j} x_{ij}$. Hence, the value of the allocation can be written as $\val(x) := \sum_{j \in J} \min(C_j, \alloc(j))$. 

\paragraph{Matching covered Graphs} Recall that a \emph{matching covered} bipartite graph is one which is connected and in which every edge is in some perfect matching. We extend the notion of matching covered to bipartite graphs that have supplies/capacities by considering perfect integral allocations: allocations $x$ such that $x_{ij} \in \Z$ for every edge $(i, j) \in E$ and so that for each $i \in I$, all supply is allocated ($\sum_{j \sim i} x_{ij} = S_i$), and for each $j \in J$ all capacity is used ($\sum_{i \sim j} x_{ij} = C_j$). Such a bipartite graph is said to be matching covered if every edge has positive allocation in some perfect integral allocation. Henceforth, we freely refer to allocations (integral or fractional) as matchings. 

An equivalent condition to being matching covered is that the graph has no nontrivial tight sets. That is, the graph does not contain a non-trivial subset $X$ of $I$ for which the set of neighbors $N(X)$ satisfies $C_{N(X)} = S_X$. Since the graph has a perfect matching, we know that Hall's condition is satisfied for every subset of $I$, and in particular $I$ is a tight set: $C_{N(I)} = S_I$. The following observation is folklore.

\begin{observation}
\label{obs:mc}
    A connected bipartite graph $G = (I \cup J, E)$ with a perfect matching is matching covered if and only if for every nonempty strict subset $X \subsetneq I$, Hall's condition is satisfied with some slack. That is, $C_{N(X)} > S_X$.
\end{observation}

This can be proved by the same method as Hall's perfect matching theorem (see \cite[\S2.3]{LM24}, for example). 

\section{Perfect Proportional Allocations}
\label{sec:ppa}
In this section, we characterize those graphs that admit a perfect proportional allocation and use this to give a new allocation strategy in those graphs that do not. We can assume that the instance is connected and has a perfect matching. That is, a matching of size $n := S_I = C_J$. Removing these assumptions is straightforward\footnote{If an instance has multiple connected components, they can be treated separately; if any instance does not have a perfect matching, it can be augmented with additional supply/capacity so that it does have a perfect matching, then these additions can be removed after assigning the weights $\alpha_j$.}. We now state the main theorem of this section. 

\begin{theorem}\label{thm:perfect-prop}
An instance of (connected) bipartite matching has a perfect proportional allocation if and only if it is matching covered. 
\end{theorem}

The theorem follows from a classical result due to Rothblum and Schneider~\cite{RS89} on so-called \emph{matrix scaling}. Given $A = (A_{ij})$ an $m \times n$ non-negative matrix, a \emph{scaling} of $A$ is a matrix $\widetilde A = (\widetilde A_{ij})$ that can be written as $\widetilde A_{ij} = A_{ij} x_iy_j$ for some positive scaling vectors $x \in \R^n, y \in \R^m$. For vectors $r \in \R^n$ and $c \in \R^m$, an $(r, c)$-scaling of $A$ is a scaling whose row and column sums are equal to $r$ and $c$ respectively. That is, a scaling $\widetilde A$ for which
\[
    \widetilde A \ind = r \qquad \text{ and } \qquad \widetilde A^T \ind = c
\]
If such a scaling exists, then we say $A$ is $(r, c)$-scalable. 

The Rothblum and Schneider matrix scaling result was restated combinatorially in the following way by Hayashi, Hirai, and Sakabe~\cite{HHS24}. For a matrix $A$, let $G = ([n] \sqcup [m], E)$ denote the bipartite graph having edge $ij$ whenever $A_{ij} \neq 0$. We then let $X \sqcup Y$ denote a set of vertices $X \subseteq [n]$ and $Y \subseteq [m]$. Let $\mc{S}$ denote the set of independent sets in $G$. 

\begin{theorem}[\cite{RS89}]\label{thm:scaling}
$A$ is $(r, c)$-scalable if and only if $\sum_i r_i = \sum_j c_j$, and $r(X) + c(Y) \leq \sum_i r_i$ for every $X \sqcup Y \in \mc{S}$, with equality only if $A[[n] \setminus X, [m] \setminus Y] = \mc{O}$, the all-zeros matrix. 
\end{theorem}

We now show how \Cref{thm:perfect-prop} follows from \Cref{thm:scaling}.

\begin{proof}
Consider the bipartite graph $G = (I \cup J, E)$, and suppose $G$ has a perfect proportional allocation given by weights $\alpha$. Let $A$ be the bipartite incidence matrix of $G$: that is, $A_{ij}$ is 1 if and only if $i \in I$ is adjacent to $j \in J$ in $G$. Now consider the scaling vectors $y = \alpha$ and $x_i = \frac{S_i}{\sum_{a \sim i} \alpha_a}$. Then we have, for each $j \in J$,
\[
    \sum_{i} A_{ij} x_i y_j = \sum_{i \sim j} \frac{S_i \cdot \alpha_j}{\sum_{a \sim i} \alpha_a} = \alloc(j) = C_j
\]
since $\alpha$ is a perfect proportional allocation. Similarly, for each $i \in I$,
\[
    \sum_{j} A_{ij} x_i y_j = \sum_{j \sim i} \frac{S_i \cdot \alpha_j}{\sum_{a \sim i} \alpha_a} = S_i.
\]
Therefore, $A$ is $(S, C)$-scalable, where $S$ and $C$ are the vectors of node supplies and capacities, respectively. By \Cref{thm:scaling}, it must be the case that $S_X + C_Y \leq n$ for every $X \sqcup Y \in \mc{S}$, with equality only if $A[I \setminus X, J \setminus Y] = \mc{O}$. In particular, for any nontrivial set $X \subsetneq I$, consider $Y = J \setminus N(X)$. Clearly $X \sqcup Y \in \mc{S}$, and moreover, by the connectivity assumption on $G$, we know that $A[I \setminus X, J \setminus Y] \neq \mc{O}$. Hence, $S_X + C_Y < n = \sum_{j} C_j$, so subtracting we get the desired Hall condition with slack:
\[
    S_X < \left(\sum_{j} C_j\right) - C_Y = C_{N(X)}.
\]
So by \Cref{obs:mc}, $G$ is matching covered.

Conversely, suppose that $G$ is matching covered, and consider any $X \sqcup Y \in \mc{S}$, so $Y \subseteq J \setminus N(X)$. First, if $A[I \setminus X, J \setminus Y] = \mc{O}$, then in particular, $A[I \setminus X, N(X)] = \mc{O}$, which by the connectivity assumption on $G$ means that either $X$ or $I \setminus X$ is empty. In this case, clearly $S_X + C_Y \leq n$. So assume otherwise that $X$ is nontrivial. But now since $G$ is matching covered, we must have 
\[
    S_X < C_{N(X)} \leq C_{J \setminus Y} = n - C_Y
\]
and so $S_X + C_Y < n$. So we may apply \Cref{thm:scaling} in the other direction to conclude that $A$ is $(S, C)$-scalable. Finally, consider the associated scaling vector $x$ and $y$. We have $S_i = \sum_{j} A_{ij} x_i y_j = x_i \sum_{j \sim i} y_j$, and hence $x_i = \frac{S_i}{\sum_{j \sim i} y_j}$ for each $i$. Now, choosing the weight vector $\alpha = y$, we clearly have a perfect proportional allocation as desired. 
\end{proof}

\subsection{Non Matching Covered Graphs}

If $G = (I \cup J, E)$ is not matching covered, we cannot get a perfect proportional allocation. However, in this subsection, we describe a simple fractional allocation strategy that achieves value equal to $\OPT$, and can still be carried out using only local information. Formally, each right hand node $j \in J$ will be assigned a weight $\alpha_j$ as before, in addition to a \emph{rank} $r_j$. The allocation of the supply of node $i$ will depend solely on the weight and rank of the neighbors of $i$. 

To achieve such an allocation strategy, we use the \emph{Dulmage–Mendelsohn decomposition}~\cite{DM58} (see also \cite[\S2.3.1]{LM24}), generalized to graphs with supplies and capacities. This is a partition of the vertices of a bipartite graph $G$ into subsets $X_1 \sqcup Y_1, X_2 \sqcup Y_2, \dots$ such that 
\begin{enumerate}[(a)]
    \item \label{prop:dm-mc} for each $k$, the induced bipartite graph on $X_k \sqcup Y_k$ is matching covered, and
    \item \label{prop:dm-order} for any $k < k'$, there are no edges $(i, j)$ in $E$ with $i \in X_k$ and $j \in Y_{k'}$. 
\end{enumerate}
The partition can be found efficiently in the following way (folklore):
\begin{enumerate}
    \item Fix a perfect integral allocation $x$ for $G$, and let $\mc{M}$ denote the edges with positive allocation. Construct a directed graph $\mc{D}$ by bi-directing each edge in $\mc{M}$, and directing each edge in $E \setminus \mc{M}$ to the right (that is, from $I$ to $J$).  

    \item Find the strongly connected components of $\mc{D}$. These correspond to a partition of the vertices of $G$. 

    \item Order the parts by any (decreasing) topological ordering of the directed acyclic graph of strongly connected components of $\mc{D}$, so that if there is an arc $(x_p,y_q)$ between two strongly connected components, then $p > q$. 
\end{enumerate}

For completeness, we include the proof that this gives a Dulmage-Mendelsohn decomposition as defined above.

\begin{proof}
We first argue that adjacent vertices $i$ and $j$ are in the same part if and only if the edge $(i, j)$ is in some perfect matching of $G$. 

If $i$ and $j$ are in the same part, then either they are matched in $\mc{M}$, or otherwise they are in the same strongly connected component in $\mc{D}$. In the latter case, there is a directed cycle in $\mc{D}$ using the arc $(i, j)$ that defines an alternating path that gives rise to a new perfect integral allocation $x'$ that assigns some of the supply of $i$ to $j$. In either case, $(i, j)$ is in some perfect matching. 

Conversely, if $(i, j)$ has positive allocation in a perfect integral allocation $x'$ using edges $\mc{M}'$, but is not in $\mc{M}$, then the symmetric difference $\mc{M} \triangle \mc{M}'$ contains an alternating cycle that includes the edge $(i, j)$. This cycle defines a directed cycle in $\mc{D}$ containing $i$ and $j$, and hence they are in the same strongly connected component. Therefore, $i$ and $j$ are in the same part. 

As a consequence, every edge contained in a part is in a perfect matching of $G$, which must also be a perfect matching on the bipartite subgraph induced by the vertices in that part. So each part is matching covered and \ref{prop:dm-mc} is true as desired. 

Finally, to see that \ref{prop:dm-order} holds, just observe that every edge between parts in $G$ corresponds to an arc between strongly connected components in $\mc{D}$, and since we ordered the parts by a topological ordering, any edge $(i, j)$ in the directed acyclic graph on the strongly connected components with $i \in X_k$ can only have $j \in Y_{k'}$ for $k > k'$. 
\end{proof}

Now we can proceed with the description of the allocation strategy. Informally, each right hand node $j$ is given a weight $\alpha_j$ and a rank $r_j$. Then nodes $i \in I$ allocate their supply proportionally among their highest ranked neighbors. 

\begin{definition}
For a bipartite graph $G = (I \cup J, E)$ with supplies and capacities, a \emph{piecewise proportional allocation} is an assignment of weight $\alpha_j \in \R_{> 0}$ and rank $r_j \in \Z$ to each node $j \in J$, with the associated allocation $x$ defined by
\[
    x_{ij} := \begin{cases}
        \frac{S_i \cdot \alpha_j}{\sum_{j' \in N_r(i)} \alpha_{j'}} & j \in N_r(i)\\
        0 & \text{otherwise}
    \end{cases}
\]
where $N_r(i)$ is defined as $\{j \in J : j \sim i \text{ and } r_j = \max_{j' \sim i} r_{j'}\}$, the set of max rank neighbors of $i$. 
\end{definition}

The value of a piecewise proportional allocation is defined identically to a proportional allocation. With the Dulmage-Mendelsohn decomposition in hand, we can prove the following theorem. 
\begin{theorem}
Every bipartite assignment instance has a perfect piecewise proportional allocation. That is, one whose value is equal to $n$. 
\end{theorem}

\begin{proof}
Consider the Dulmage-Mendelsohn decomposition $X_1 \sqcup Y_1, X_2 \sqcup Y_2, \dots$ of $G$. Assign for each $j \in J$ a rank $r_j$ equal to the index of the part to which it belongs: $Y_{r_j}$. For each part $X_k \sqcup Y_k$, find the perfect proportional allocation on the subgraph induced by this part, and give each $j \in Y_k$ its corresponding weight $\alpha_j$ from that proportional allocation. 

The proof follows almost immediately from the definition of the Dulmage-Mendelsohn decomposition: since each $X_k \sqcup Y_k$ is matching covered by property \ref{prop:dm-mc}, \Cref{thm:perfect-prop} implies that it has a perfect proportional allocation $x^{(k)}$ with corresponding weights $\alpha^{(k)}$. Moreover, property \ref{prop:dm-order} implies that each $i \in X_k$ has edges only to $j \in Y_{k'}$ with $k' \leq k$. In particular, 
\[
    N_r(i) = N(i) \cap Y_k,
\]
and thus the piecewise proportional allocation $x$ defined by the weights and ranks above exactly agrees with $x^{(k)}$, for every $k$. Hence, the value of $x$ is the sum of values of the $x^{(k)}$, which is $n$, as desired. 
\end{proof}

\noindent{\bf Acknowledgment:} We thank Mohit Singh for pointing us to the work on matrix scaling.

\bibliographystyle{alpha}
\bibliography{bibliography}

\end{document}

%% file: preamble.tex
\usepackage[shortlabels]{enumitem}
\usepackage{amsmath} 
\usepackage{amssymb} 
\usepackage{mathrsfs} 
\usepackage{mathtools} 
\usepackage{amsthm}
\usepackage{thmtools}

\usepackage{appendix}
\usepackage{fullpage}
\usepackage{caption}
\usepackage{subcaption}
\usepackage{graphicx}
\usepackage[table,xcdraw]{xcolor}
\usepackage{wrapfig}
\usepackage{floatrow}
\usepackage{multirow}
\usepackage{makecell}

\definecolor{ForestGreen}{rgb}{0.0333,0.4451,0.0333}
\definecolor{DarkRed}{rgb}{0.65,0,0}
\definecolor{Red}{rgb}{1,0,0}
\usepackage[linktocpage=true,
pagebackref=true,colorlinks,
linkcolor=DarkRed,citecolor=ForestGreen,
bookmarks,bookmarksopen,bookmarksnumbered]{hyperref}

\usepackage[ruled,vlined,linesnumbered]{algorithm2e}
\definecolor{forestgreen}{rgb}{0.13, 0.55, 0.13}

\SetCommentSty{mycommfont}

\usepackage{cleveref}
\usepackage{thm-restate}
\usepackage[normalem]{ulem} 
\usepackage{nicefrac}

\newtheorem{theorem}{Theorem}[section]

\newtheorem{observation}[theorem]{Observation}

\theoremstyle{definition}
\newtheorem{definition}[theorem]{Definition}
\theoremstyle{remark}

\newcommand*\mb[1]{\mathbb{#1}}

\newcommand*\mc[1]{\mathcal{#1}}
\DeclarePairedDelimiter\abs{\lvert}{\rvert}
\DeclarePairedDelimiter\norm{\lVert}{\rVert}%

\makeatletter
\let\oldabs\abs
\def\abs{\@ifstar{\oldabs}{\oldabs*}}
\let\oldnorm\norm
\def\norm{\@ifstar{\oldnorm}{\oldnorm*}}
\makeatother

\newcommand{\eps}{\varepsilon}

\newcommand{\OPT}{\text{OPT}}

\newcommand{\R}{\mb{R}}

\DeclareMathAlphabet{\mymathbb}{U}{bbold}{m}{n}
\newcommand*\mmb[1]{\mymathbb{#1}}
\newcommand*\ind{\mmb{1}}

\DeclareMathOperator{\val}{value}
\DeclareMathOperator{\alloc}{alloc}

\newcommand{\Z}{\mb{Z}}

\newif\ifcomments
\commentstrue
\commentsfalse 

\ifcomments
\usepackage[colorinlistoftodos,prependcaption,textsize=tiny]{todonotes}

\definecolor{dnotecol}{rgb}{0.20, 0.50, 0.80}

\else 

\fi 